\documentclass[preprint,secnumarabic,amssymb, nobibnotes,aps,prd]{revtex4-2} % twocolumn

\newcommand{\nn}{\nonumber}
\usepackage{graphicx,epsfig,wrapfig}
\usepackage{float}
\usepackage{mathtools}
\begin{document}
\title{Optimization of second-harmonic generation from touching plasmonic wires}

\author{Shimon~Elkabetz$^1$}
\email{elshimon@post.bgu.ac.il}
\author{K.~Nireekshan~Reddy$^1$}
\email{kothakap@post.bgu.ac.il}
\author{Parry~Y.~Chen$^1$}
\author{Antonio~I.~Fern\'andez-Dom\'inguez$^2$}
\author{Yonatan~Sivan$^1$}
\affiliation{$^1$ School of Electrical and Computer Engineering, Ben-Gurion University of the Negev, Beer-Sheva, 8410501, Israel}
\affiliation{$^2$ Departamento de F\'isica Te\'orica de la Materia Condensada and Condensed Matter Physics Center (IFIMAC), Universidad Aut\'onoma de Madrid, E-28049 Madrid, Spain}
\date{\today}%
% We compare between a simple single wire structure and the TWs for background permittivity equal to 12. We observe 5 orders of magnitude increase of efficiency with respect to a single wire. Surprisingly, the near field response of the TWs increase in 2 orders of magnitude and the far field demonstrate increase of scattered power by 10 orders of magnitude by increasing the background permittivity to 12.
% send to Paloma, Fan Yang and Yu Luo.
\begin{abstract} 
We employ transformation optics to optimize the generic nonlinear wave interaction of second-harmonic generation from a pair of touching metallic wires. We demonstrate a 10 orders of magnitude increase in the second-harmonic scattering cross-section by increasing the background permittivity and a 5 orders of magnitude increase in efficiency with respect to a single wire. These results have clear implications for the design of nanostructured metallic frequency-conversion devices. Finally, we exploit our analytic solution of a non-trivial nanophotonic geometry as a platform for performing a critical comparison of the strengths, weaknesses and validity of other prevailing theoretical approaches previously employed for nonlinear wave interactions at the nanoscale.
\end{abstract} 
\maketitle

\section{Introduction}

The advent of lasers has opened the gateway to studying and understanding various nonlinear optical effects and to harnessing them for practical purposes. The simplest of these effects are the second-order nonlinear processes, and in particular, their degenerate version, second-harmonic generation (SHG). % are symmetry forbidden in the local bulk response limit, and thus are governed by higher-order non-local bulk and surface (due to broken inversion symmetry at the interfaces) effects \cite{}. 
Using standard nonlinear optical materials, commercial devices based on SHG are already available (e.g., a green laser pointer). Nevertheless, recent advances in fabrication technology may pave the path towards further increases of device efficiency via nanostructuring. Of particular interest in this context are metal-dielectric nanostructures that benefit from the large field intensities associated with plasmonic resonances.

% single metallic nanoparticles~\cite{}, split-ring resonators [17,18], periodic nanostructured metal films [19], and other nano-geometries [see Ref. 20, and references therein for a more thorough review]. % The theoretical modeling of nonlinear response from these metallo-dielectric interfaces relied either on surface nonlinearity \cite{} or non-local bulk nonlinearity \cite{}, while some studies have incorporated both effects \cite{}. 

Theoretical research into this problem has focused upon two somewhat distinct aspects. First, attention has been devoted to understanding the correct description of the underlying optical response of the metal itself, from standard electronic-like second-order polarization tensors or even simpler surface polarization tensors~\cite{Kauranen_perp_perp_perp} to hydrodynamic models with growing complexity, see~\cite{Bloembergen:1968vc,scalora_model_SHG_THG}. The latter set of studies included the identification of the relative importance of the various terms in the model~\cite{jnl,Ciraci:2012_PRBR,Ciraci:2012vw,Lienau_hydrodynamic} and its specialization to even more complex spatial configurations such as hetero-dimers~\cite{Niv_SHG_2018,Niv-Schvartzman-Sivan}. Second, efforts have been dedicated to understanding the wave physics associated with the various structures of interest (see e.g.,~\cite{nl_scattering_theory_Bonn,Bachelier:2008vm,Butet:2010es,Butet_nl_Mie,Martin_spheres,Pasha_modal_coupling}), connecting it to the linear response using simple models~\cite{Miller_rule,nl_scattering_theory_Bonn,Haim_overlap}, understanding the underlying symmetries~\cite{symmetry_break_Zyss,Salomon_Zyss_triangles} and optimizing the response (see e.g.,~\cite{Taiwanese_double_resonance_films,Italians_double_resonance}).

% somewhere worth saying that we focus on particles and single ones... rather than arrays and/or wg geometries.

%As for the linear response of complex metallic nanostructures, this class of problems enjoys a full analytic description only for a scarce number of simple geometries.
As for the linear response of complex metallic nanostructures, fully analytical solutions are available for only a limited number of simple geometries. Instead, previous theoretical work has mostly been numeric, and usually addresses only the far-field properties of the SH fields~\cite{Martin_spheres}. In~\cite{Reddy_SHG_TO}, we introduced a new analytic tool for this class of problems, namely, Transformation Optics (TO). This technique rose to fame in facilitating the design of the invisibility cloak~\cite{Pendry_science_cloaking}, and was later used for the interpretation and design of a range of plasmonic structures~\cite{TO_review_Alex_2012,FJ_trans_plasmonic,Antonio_TO_review,Antonio_Paloma_TO_review}. Our implementation is seemingly the first to use TO to study nonlinear optical wave mixing. Specifically, we analytically calculated the second harmonic (SH) field generated by a pair of identical touching metal wires (TWs); see complete details in Appendix~\ref{app:PRB}. % A schematic of the the solution procedure is shown in Fig.~\ref{fig:TWs}, and its details are provided in Appendix~\ref{app:PRB}.

% can you also please help in adding references to as many leading groups - Luis Martin's, (Bouhelier?, , ... any relevant american (shanhui Fan? - wg.. metioned for THz), australian and chinese groups - wg... }

% Specifically, we used TO to provide an analytic solution to the problem of second harmonic (SH) generation from two identical touching wires. Exploiting previous studies of the linear response~\cite{Alex_kissing_cyls_NL,alex_kissing_NJP,Antonio_kissing_spheres_PRL}, we identified the various physical considerations affecting the efficiency of the SH generation and demonstrated excellent agreement to the numerical solution.

\begin{figure}[ht]
\centering
\includegraphics[width=1\linewidth]{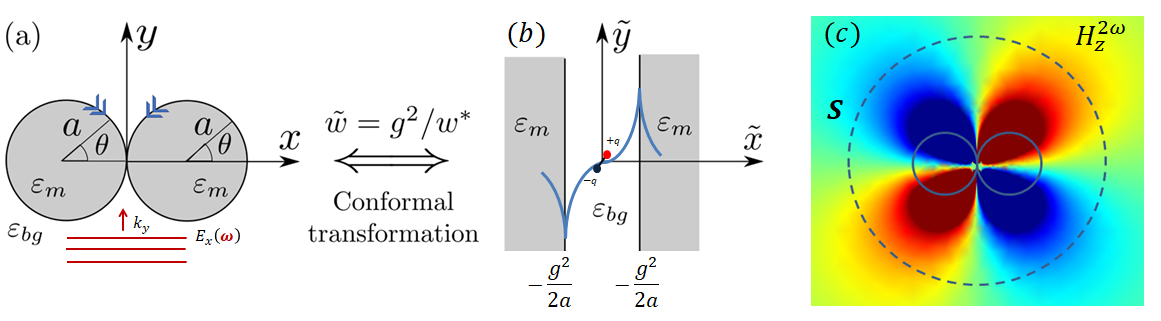}
\caption{(Color online) Schematics of (a) the (identical) touching wire system and (b) the slab geometry to which it is related via an inversion transformation. (c) The scattered second harmonic magnetic field; a quadrupolar pattern with distinct anti-symmetry in both $x$ and $y$ is observed. %$\hat{n}$ represents the normal of a closed contour $\mathcal{S}$ that surrounds the TWs. } 
$\mathcal{S}$ represents a closed contour that surrounds the TWs. }\label{fig:TWs}
\end{figure}

The Transformation Optics approach enabled several qualitative insights. First, it provided a simple interpretation of the analytic solution~(\ref{eq:ShMagneticField}), outlined below. The fundamental frequency (FF) plane wave which is incident upon the TWs (see Fig.~\ref{fig:TWs}(a)) transforms to a point source in the slab frame~\cite{Alex_kissing_cyls_NL,alex_kissing_NJP} (Fig.~\ref{fig:TWs}(b)), exciting a slab mode. The resulting SH source can be calculated and transformed back to the TW frame, as discussed in detail in~\cite{Kauranen_perp_perp_perp,Reddy_Sivan_SHG_BCs,Reddy_SHG_TO}. It has the generic form
\begin{equation}
J_{z,r}(x,y) = \frac{\chi^{(2)}_{S,\perp\perp\perp}}{\varepsilon_{bg}^{2\omega}}~\partial_\parallel (E_\perp^\omega E_\perp^\omega)~\delta(x^2 + y^2 - 2ax), 
\label{eq:jmz}
\end{equation}
where $x$ and $y$ are the spatial coordinates (see Fig.~\ref{fig:TWs}(a)) $\chi^{(2)}_S$ is the surface second-order susceptibility perpendicular to the interface, $E_\perp^\omega$ is the FF electric field perpendicular to the interface, and $\varepsilon$ is the permittivity. 

%When this solution is transformed back to the TW frame, it becomes an anti-symmetric quadrupolar profile, see Fig.~\ref{fig:TWs}(c). 
This solution reveals why the SH field profile is quadrupolar and anti-symmetric, see Fig.~\ref{fig:TWs}(c). The FF modes excited in the slab frame have electric fields that are symmetric in $x$ and $y$, since these are the surface plasmon waves that exist within the frequency band below the plasma resonance frequency (see~\cite{Reddy_SHG_TO}). The square of the FF electric field in \eqref{eq:jmz} is symmetric, but the differentiation operator %this FF electric source excites surface plasmon waves only within the frequency band below the plasma resonance frequency (see~\cite{Reddy_SHG_TO}).%~\footnote{Specifically, $J_{z,r}(x,y) = - J_{z,l}(-x,y)$ and $J_{z,r}(x,y) = - J_{z,l}(x,-y)$.}
causes the source to be antisymmetric in $x$ and $y$, so it excites the higher frequency asymmetric slab mode via the so-called mode matching (MM) condition (implicit in Eq.~(\ref{eq:ShMagneticField}), see~\cite{Reddy_SHG_TO}), see Fig.~\ref{fig:TWs}(b). Transforming back to the TW frame yields the observed profile. % Nireekshan's plot from the end shows that the PM factor is smaller for 1... so maybe PM is dominated by the imaginary part...
 % ; this observation % was overlooked in~\cite{Reddy_SHG_TO}, and 
% is inline with the so-called selection rules~\cite{Dadap:1999uc,Reddy_SHG_TO}. 

Second, the TO solution~(\ref{eq:ShMagneticField}) revealed that the magnitude of the SH field is proportional to three factors. The first is the strength of the SH source, $J_{z,r}$~(\ref{eq:jmz}). The second is the magnitude of the so-called phase matching (PM) parameter 
\begin{equation}%\label{eq:P}
\mathcal{P}(\omega) = \cosh \left(2\alpha^\omega\right) + \frac{\varepsilon^{2\omega}_{bg}}{\varepsilon^{2\omega}_m} \sinh \left(2\alpha^\omega\right),
\label{eq:phasematching} 
\end{equation} 
where $\alpha^\omega$ is the propagation constant as well as the mode transverse width. This represents the pole of the dispersion relation of the antisymmetric slab mode, i.e., the proximity to the SH resonance of that mode. Third, the SH magnetic field is also proportional to a newly identified geometric factor, see~(\ref{eq:GF}) which suppresses SHG at the touching point and originates from the non-uniform distributed nature of the source across the metal-dielectric interface. %The symbols $\chi^{(2)}_S$, $E_\perp^\omega$, $\varepsilon$ above are the perpendicular surface second-order susceptibility, the perpendicular FF electric field and permittivity, while $\alpha^\omega$ represents the propagation constant as well as the mode transverse width. 
Further details on the physical interpretation of these terms and explanations of the parameters on which they depend are supplied in Appendix~\ref{app:PRB}.

% note that we do not know where are the higher order multipoles - I speculated that they are the photonic modes (Fabry-Perot modes, those above the light line). these were neglected in the initial linear calculations (branch cut), and later accounted for as lossy surface waves (localized in y, confined in x). but in KNR's calculation, which is somehow beyond the QS, they might appear...? related to hidden symmetries?}

%\begin{figure}[ht]
%\centering
%\includegraphics[width=0.7\linewidth]{QPscattredfield2.png}
%\caption{(Color online) (a) Two surfaces $\mathcal{S}_1$ and $\mathcal{S}_2$ surrounding the TWs. $\tau_x$ and $\tau_y$ map each point on these surfaces to the TWs perimeter. For $\mathcal{S}_1$%, with a radius of 22nm
%, each point maps to points relatively far from the touching point. Conversely, for $\mathcal{S}_2$ which is closer to the TWs structure, each point maps closer to the touching point. (b) The scattered second harmonic magnetic field; a distinct anti-symmetric pattern in $x$ and $y$ is observed, which corresponds to a quadrupolar response. $\hat{n}$ represent the normal of the surrounding surface, $\mathcal{S}$.} \label{ScatterdColorMap}
%\end{figure}

In the present paper, we proceed beyond~\cite{Reddy_SHG_TO} by exploiting the analytic solution~(\ref{eq:ShMagneticField}) and the qualitative insights it provides to perform a {\em quantitative} study of the SHG from the TWs. Specifically, we study the interplay between the three aforementioned factors to optimize the near-field enhancement, as well as the far-field response, as quantified by the scattered power at the SH wavelength, $P^{2\omega}_{scat}$. The paper is organized as follows. In Section~\ref{sec:config}, we describe the configuration under study and the methodology. In Section~\ref{sec:1vs2}, we compare the SH response of the TWs with that of a single wire, demonstrating up to 5 orders of magnitude greater SHG efficiency. In Section~\ref{sec:bg_epsilon}, we explore how the SHG of the TWs varies with $\varepsilon_{bg}$ and show that the scattered power can be increased by up to 10 orders of magnitude by increasing $\varepsilon_{bg}$. In Section~\ref{sec:spectrum}, we analyze the spectral response and identify the origins of its narrowband nature. In the last section of the paper, we exploit our analytic solution of a non-trivial nanophotonic geometry as a platform to perform a critical comparison of the strengths, weaknesses and validity of other prevailing theoretical approaches employed for nonlinear wave interactions at the nanoscale. Finally, we conclude with an outlook.

%and show that its increase can lead to an increase of the scattered power by up to 10 orders of magnitude.

\section{Configuration and methodology}\label{sec:config}
For all results presented in this manuscript, we choose a wire radius of $a = 5$ nm, with a permittivity characterized by a Drude model (with parameters suitable for Ag, namely, $\varepsilon_\infty = 5$, $\omega_p = 9.2$eV and $\gamma = 0.2$eV~\footnote{Note that as in~\cite{Reddy_SHG_TO}, we chose a relatively high value for the imaginary part of the metal permittivity, corresponding to low quality metal nanostructures which are appropriate for several fabrication techniques; this choice was made in order to ensure convergence of the numerical simulations. }). We also choose $\chi^{(2)}_{S,\perp\perp\perp} = 10^{-20}$ m$^2$/V in all simulations. However, all SH results in this paper are linearly proportional to this parameter, so it can be adjusted to suit any particular nonlinear polarization model. We compare two dissipationless background permittivities: low ($ = 1$, considered previously in~\cite{Reddy_SHG_TO}) and high ($ = 12$, characterizing semiconductor materials with negligible losses, i.e., when illuminated at a frequency lower than the energy bandgap of the semiconductor). For simplicity, we also assume the host to be dispersionless (i.e., $\varepsilon_{bg}^{2\omega} = \varepsilon_{bg}^\omega = \varepsilon_{bg}$). To ensure a fair comparison, we choose the incident FF field to be an $x$-polarized (i.e., along the dimer axis) plane wave with unit amplitude in both cases. % As shall be shown, these represent two extremes of the conversion efficiency. {\bf does it?}

We discuss below three main types of results. % ({\bf I think the sections should be precisely specified.}) 
We first focus on the near-field SH wave along the perimeter of the wire close to the touching point, marked by $C$. Due to symmetry, we consider only the right wire of the TW set. The $H_z^{2\omega}$ field is given by (Appendix~\ref{app:PRB})
\begin{equation}
H_z^{2\omega}(x,y)|_C = - i \omega \varepsilon_0 \varepsilon_{bg}^{2\omega} \frac{ \sinh 2\alpha^\omega}{\alpha^\omega \mathcal{P}} ~x~ J_{z,r}(x,y). \label{sh_sol_simp}
\end{equation}
%({\bf The notation of this equation and Appendix A are inconsistent, since P appears here and R appears there.}) 
This demonstrates that the resulting $H_z^{2\omega}$ at each point is proportional to $J_{z,r}$ at that point, subject to the two aforementioned additional factors. These are the PM parameter defined in~\eqref{eq:phasematching}, and the geometric factor~(\ref{eq:GF}), which has been reduced to $x$ along the TW perimeter. %, which suppresses the effect of $J_{z,r}$ near the origin. 
% ({\bf I think its better to consolidate this discussion with (2) and the text that surrounds it.})
%This shows that along the TW perimeter, the product of the geometric factor~(\ref{eq:GF}) and $\mathcal{R}$ (see Eq.~(\ref{eq:Decpomposedjmz})) reduces to the product of the SH source $J_{z,r}$ and $x$ {\bf Parry - would you suggest to say anything further on the geometric factor here?}.

We also present the scattering cross-section; the details of its analytic and numeric calculations are given in Appendix~\ref{app:cs}. Finally, in the context of the spectral response, we also present various % various quantities of the solution for the two cases above - the maximal {\bf SE - we did not do that, did we?\textcolor{blue}{Only the average values of the near-field magnetic response over the TWs perimeter. }} and 
averaged values of near-field properties %(the $L_\infty$ and $L_1$ norms, respectively) 
over the TW perimeter, defined as
\begin{eqnarray}
\langle |f(x,y)|\rangle_C = \frac{1}{4\pi a}\int_C |f(x,y)|\, dx dy.
\end{eqnarray}

As observed in~\cite{Reddy_SHG_TO}, the source $J_{z,r}$~(\ref{eq:jmz}) and magnetic field~(\ref{sh_sol_simp}) were found to be in excellent agreement with the numeric solution only relatively close to the touching point, and not too close to PM% ({\bf the PM condition?})
. Therefore, whenever we discuss the solution far away from the touching point, we employ the numeric solution instead of the analytic one (e.g., for cross-section calculations). As is customary, analytic results are marked by lines within figures, and numerical results by symbols.

\section{Single wire vs. the touching wires}\label{sec:1vs2}

\begin{figure*}
\centering
\includegraphics[width=1\linewidth]{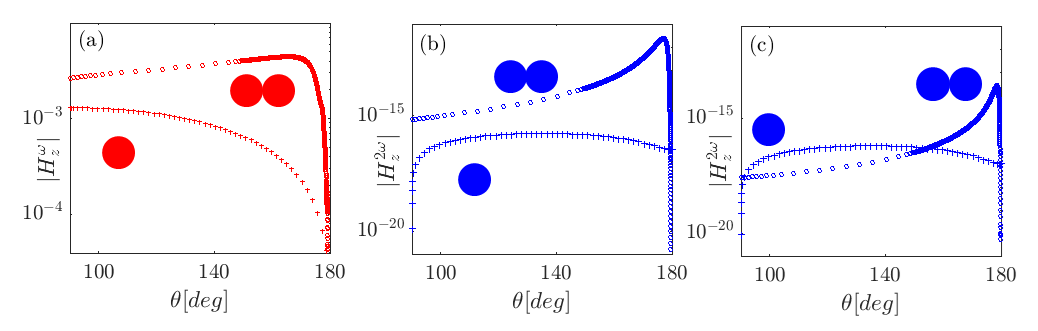}
\caption{(Color online) (a) The numeric solution for the linear magnetic field $H_z^\omega$ along the perimeter of the TWs ({\texttt{$\circ$}}) and the single wire ({\texttt{+}}) as a function of angle $\theta$ for $\varepsilon_{bg} = 12$ at a fundamental wavelength of $922$nm (hence, $\lambda_{SH} = 461$ nm). % ({\bf why do fields of the single cylinder change with angle? SE - is this just a trigonometric dependence? write down the dependence on theta from your solution. ANSWER: The solution is similar to a dipole scattering. And for the near fields it seems like a dependence on a cosine} 
(b) Same as (a) for the SH magnetic field $H_z^{2\omega}$. (c) Same as (b) at a fundamental wavelength of $1400$nm. The case of $\varepsilon_{bg} = 1$ is qualitatively similar (data not shown). In all cases, the incident electric field is an x-polarized plane wave with unit amplitude. }
\label{nrfld_TW_comp} 
\end{figure*}

% {\bf KNR/SE - is there something like the $G$-factor for the single wire? 
As an initial step, we compare the SH responses of the touching wires and a single wire. Fig.~\ref{nrfld_TW_comp}(b) shows that the TWs generate stronger SH near-fields for virtually all incidence angles due to the intrinsic geometric singularity in the TWs. Both structures are near resonance.  Similar behaviour is observed in the linear case (i.e., at the FF), as seen in Fig.~\ref{nrfld_TW_comp}(a).
% but also oscillates more rapidly close to touching point. 
Fig.~\ref{nrfld_TW_comp}(c) demonstrates the situation further from resonance, corresponding to a longer wavelength. Here, the TWs generate a stronger near-field close to the touching point only. 

\begin{figure}[H]
\centering
\includegraphics[width=1\linewidth]{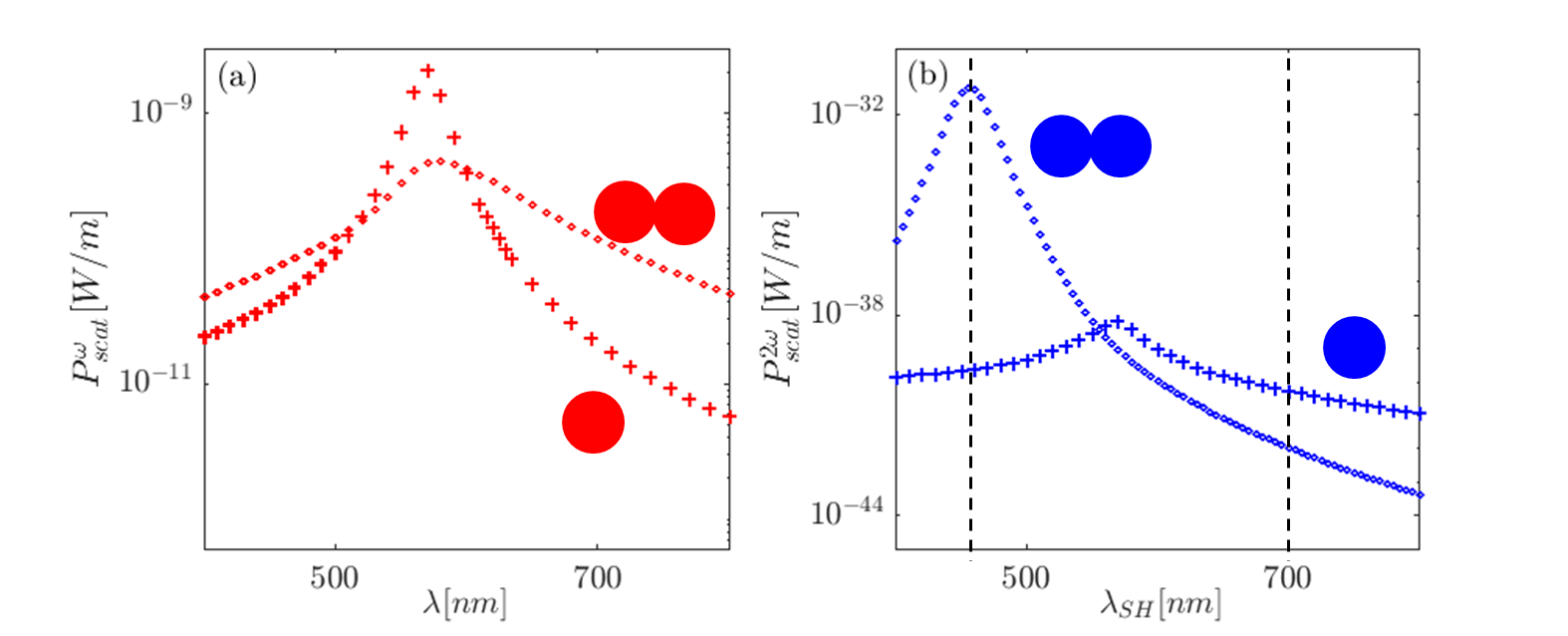}
\caption{% Same as Fig.~\ref{TWsVsSingle} for $\varepsilon_{bg} = 12$. 
(Color online) (a) The numeric solution for the linear (i.e., at the FF) scattered power of TWs (\texttt{+}) and single wire (\texttt{$\circ$}). (b) The corresponding solution for the SH scattered power; the black dashed lines represent the scattered power for fundamental wavelengths $922$nm and $1400$nm, corresponding to the near-field plots in Fig.~\ref{nrfld_TW_comp}. Both scattered powers are calculated for $\varepsilon_{bg} = 12$. % The results for $\varepsilon_{bg} = 1$ are qualitatively similar (not shown) {\bf I actually want to see it - is there a spectral regime where the TWs perform better than the single wire?\textcolor{blue}{See \url{mono-dimer-cs-comparison_1bg}}}. 
} \label{fig:TWsVsSingle12}
\end{figure}

The behaviour of the scattered power is somewhat more complicated. Fig.~\ref{fig:TWsVsSingle12}(a) shows that single wire exhibits stronger linear scattering compared with the TWs when both are at resonance~\footnote{Peculiarly, since previous studies of this problem focused on a comparison of the spectral bandwidth of these two structures, this result was not noted previously. }. However, since the spectral bandwidth of the TW response is much wider~\cite{Alex_kissing_cyls_NL,alex_kissing_NJP}, the TW response is stronger away from resonance. Surprisingly, the situation is inverted for the SH scattered power, as seen in Fig.~\ref{fig:TWsVsSingle12}(b). In particular, the power scattered at resonance by the TWs is greater by up to 5 orders of magnitude! Yet, since its spectral width turns out to be significantly narrower, the power scattered at longer wavelengths is greater for the single wire. This is the first main result of the current work. This behaviour can be traced to the relatively weak SH fields generated at points along the TW perimeter further from the touching point in the off-resonance case, see Fig.~\ref{nrfld_TW_comp}(c), as it is these points that determine the far-field scattered power (see Fig.~\ref{CS_1_NumVsAn}(a)), 

Also worth noting is that the peaks of FF and SH scattered powers are at the same wavelength for the single nanowire, and these coincide with the peak of the TW linear response. However, the peak SH response of the TW is blue-shifted. The origin of this blue-shifting can be understood from the modal analysis enabled by TO. In particular, all plasmonic modes lie exactly at the surface plasmon frequency for the single nanowire in the quasi-static limit; indeed, this configuration corresponds to a single interface plasmonic waveguide in the transformed slab frame. This corresponds to $\sim 580$ nm (see Fig.~\ref{fig:TWsVsSingle12}) for which ${\rm Re}[\epsilon_m] \approx -12$. Meanwhile, two branches emerge in the dispersion relation for the TWs (which transform to the slab waveguide, Fig.~\ref{fig:TWs}(b)), one above and one below the surface plasmon frequency. The upper branch corresponds to the antisymmetric modes, and is characterized by a negative group velocity. As a result, the modes corresponding to smaller wavevector are those with highest frequency and the wider spatial extent; the latter corresponds to the highest radiative damping. A more detailed discussion and as well as a visualization can be found in~\cite{Antonio_ACS_Photonics_2018}. Since the SH field is asymmetric (see Fig.~\ref{fig:TWs}(c)), the SH peak of the TWs occurs at frequencies that correspond to the higher frequency branch, above the surface plasmon frequency and approaching the metal plasma frequency.

% \begin{figure}[H]
% \centering
% \includegraphics[width=1\textwidth]{LinearCSandSHCS4_.png}
% \caption{(a) The numeric solution for the normalized linear (i.e., at the FF) scattering cross-sections of TWs (pluses) and single wire (dots). (b) The corresponding solution for the SH scattering cross-sections. The blue solid line shows the analytic calculation of SH scattering cross-section. The TWs has $5$nm radius and the single wire has $10$nm radius. Both scattering cross-sections are calculated for $\varepsilon_{bg} = 1$ and normalized by $A$ which represents the square root of each structure area, $17.7$nm and $12.5$nm for the single wire and TWs, respectively. The results for $\varepsilon_{bg} = 12$ are qualitatively similar, see Fig.~\ref{fig:TWsVsSingle12}. } \label{TWsVsSingle}
% \end{figure}

% LET US REMOVE THIS PARAGRAPH... IT IS NOT SO IMPORTANT EVENTUALLY... More generally, one should recall that any realistic application would involve an array of particles. The SHG of an array of TWs is expected to be even stronger than an array of cylindrical wires, since an ordered array of the latter has inversion symmetry, and is not expected to yield a macroscopic SH response. On the other hand, an array of touching wires has reduced symmetry, hence, is expected to yield a non-zero SHG. ({\bf This depends on orientation. An array of aligned TWs can have inversion symmetry.})% In that sense, the revealed sensitivity of the SHG efficiency to the background permittivity provides a valuable design rule.

\section{Optimization of the SH response of the touching wire}\label{sec:bg_epsilon}
The analytic solutions for the near- and far-fields now allow us to scan the parameter space in search of the optimal response, by varying $\omega$, $\varepsilon_{bg}$ etc. While this has clear practical value, it is also tedious, and is worth doing only for a well-defined experimental set up. Instead, we opt to explain these numerical findings qualitatively, and to identify the governing physics that enable further optimization of the SH response. In particular, we exploit the solution~(\ref{eq:ShMagneticField}) and analyze the contributions of the previously mentioned factors to the SH response. A link exists between the near-field and far-field behaviour realized by $\mathcal{R}$~(\ref{eq:Decpomposedjmz}), see Fig.~\ref{CS_1_NumVsAn}(a), so it suffices to analyze the near-field solution~(\ref{sh_sol_simp}).

\subsection{Near-field response}
\label{sub:near}
\begin{figure}[H]
\begin{center}
\includegraphics[width=1\linewidth]{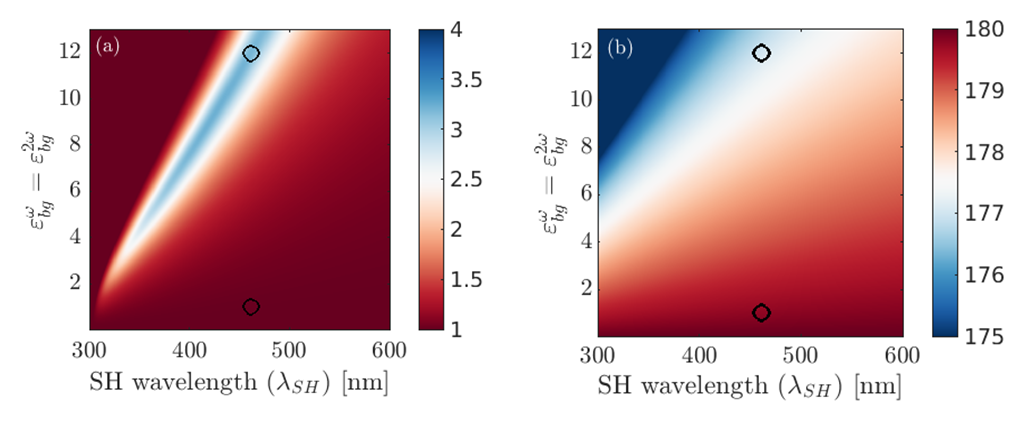}
\caption{(Color online) (a) $1/|\mathcal{P}|$~(\ref{eq:phasematching}) as a function of SH wavelength and dispersionless dielectric background $\varepsilon_{bg}$. The points marked `$\bigcirc$' denotes simulations at the SH wavelength of 461 nm, for $\varepsilon_{bg} = 12$ and $\varepsilon_{bg} = 1$. (b) $\theta_{max}^\omega$ (see Eq.~\eqref{eq:th_max}) as a function of SH wavelength and dielectric background $\varepsilon_{bg}$. }
\label{fig_poles_soln}
\end{center}
\end{figure}

We now analyze the SH generation efficiency by evaluating each of the various elements in Eq.~(\ref{sh_sol_simp}) separately. % Despite the slightly decreasing accuracy of Eq.~(\ref{sh_sol_simp}) further from the touching point, we used the analytical sol. for the cross-section part. in other places, we use the numerical results. 
We begin by finding the optimal material parameters for which the phase-matching factor $1/|\mathcal{P}|$ is maximized~\footnote{Ideally, one desires to have the optimal parameters such that $1/|\mathcal{P}|$ diverges; in this case, the geometric factor $x$ will have little consequence. However, as we consider a lossy metal, $1/|\mathcal{P}|$ attains its maximum value instead of diverging.}, i.e., the background permittivity and frequency that brings the system as close as possible to phase matching (or equivalently, to SH resonance). From Fig.~\ref{fig_poles_soln}(a), it is evident that the background permittivity required to maximize $1/|\mathcal{P}|$ is high, especially as the fundamental wavelength gets longer. For example, for a fundamental wavelength of $922$nm (hence, $\lambda_{SH} = 461$nm), the optimal choice for the background dielectric is $\approx 12$ (marked by `$\bigcirc$' in Fig.~\ref{fig_poles_soln}(a)); this motivates a-posteriori the choice of parameters in the previous section. Fig.~\ref{fig_poles_soln}(a) also shows an overall moderate sensitivity of $1/|\mathcal{P}|$ to the choice of permittivities and SH wavelength. In particular, the high permittivity case yields a maximum value of $1/|\mathcal{P}| \approx 3.2$ for $\varepsilon_{bg} = 12$, whereas $1/|\mathcal{P}| \approx 1$ for the $\varepsilon_{bg} = 1$ case. 

\begin{figure}[ht]
\begin{center}
\includegraphics[width=1\linewidth]{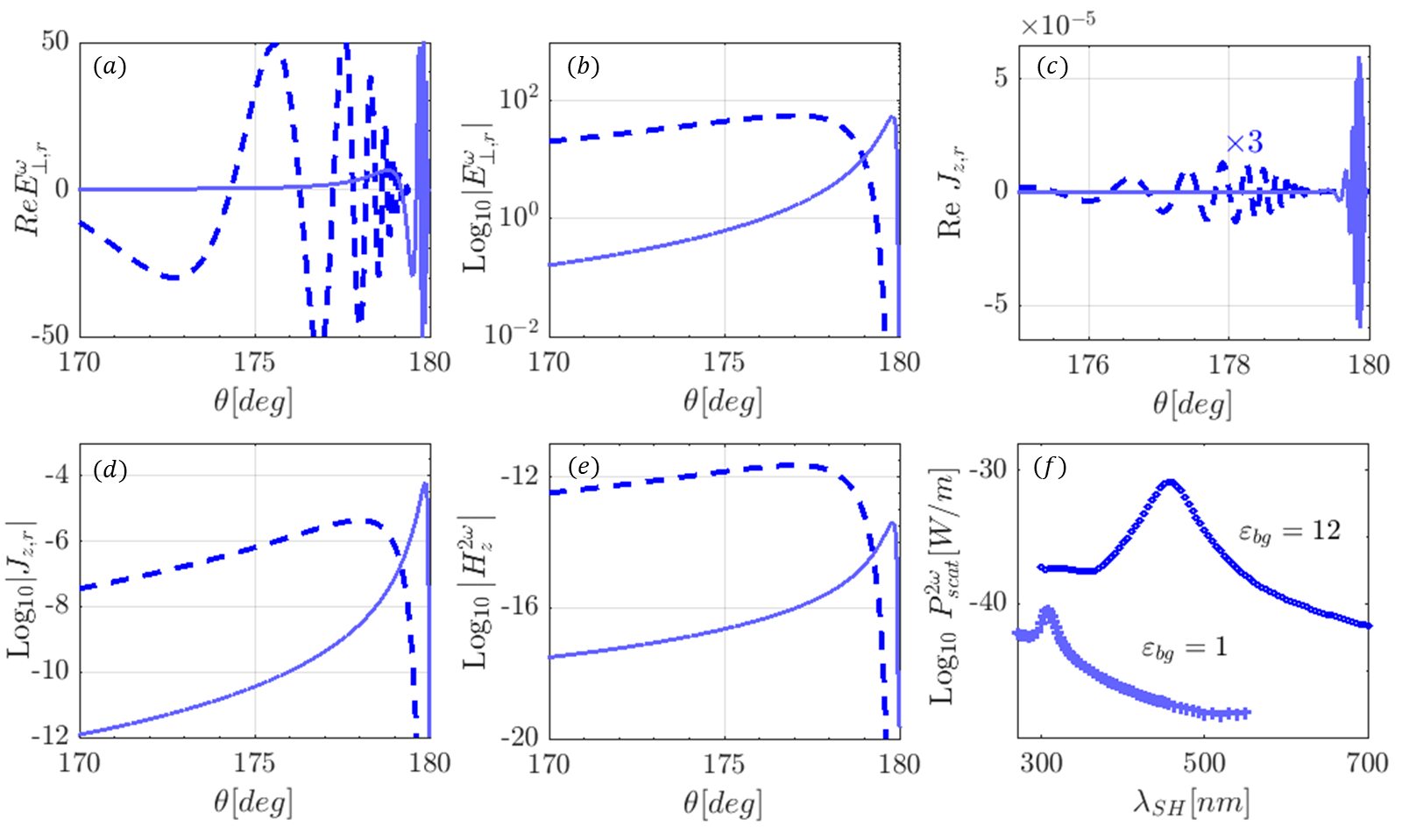}
\caption{(Color online) (a) - (e) Analytic solutions of near-fields. Fields are plotted as function of angle $\theta$ (see Fig.~\ref{fig:TWs}) along the perimeter of the right wire for the TW system%({\bf This should appear already in Fig. 2}) - it kinda did.. but there it was the H field, and the comparison made was different.
. Curves for $\varepsilon_{bg} = 1$ and $\varepsilon_{bg} = 12$ are plotted using light blue solid lines and dark blue dashed lines, respectively, and in slightly different colours. Note the particularly rapid oscillations in the former case. Linear fields (a) $Re[E_\perp^\omega]$, (b) $|E_\perp^\omega|$ at the FF wavelength $922$nm. The SH source (c) $Re[J_{z,r}]$ and (d) $|J_{z,r}|$ at the SH wavelength of $461$ nm. (e) The SH magnetic field $H_z^{2\omega}$. (f) The numeric solution for the SH scattered power of TWs for $\varepsilon_{bg} = 12$ (dots) and $\varepsilon_{bg} = 1$ (crosses). }
\label{fig:sh_diel} 
\end{center}
\end{figure}

We now turn to study the effect of $\varepsilon_{bg}$ on the product $x J_{z,r}$, which includes the geometric factor $x$. % In order to extract the SH field enhancement apart from the PM factor, we compare the SH source $J_{z,r}$ strengths.  Since $J_{z,r}$ is evaluated from the FF fields, 
To do so, we first display in Figs.~\ref{fig:sh_diel}(a) and~\ref{fig:sh_diel}(b) the real and absolute value of the FF fields $E_\perp^\omega$ along the wire perimeter for both choices of $\varepsilon_{bg}$. The maxima of the fields are roughly of the same order of magnitude for both cases. While both maxima are located close to the touching point, the maximum for $\varepsilon_{bg} = 12$ is further from the touching point, corresponding to smaller angles. Indeed, it was shown in~\cite{alex_kissing_NJP} that the position of the peak value of the FF fields can be approximated by 
\begin{eqnarray}
\theta_{max}^\omega \approxeq \pi - \frac{\textrm{Im~} \left(\frac{\varepsilon_m^\omega}{\varepsilon_{bg}} \right)}{\left|\frac{\varepsilon_m^\omega}{\varepsilon_{bg}}\right|^2 - 1}, \label{eq:th_max}
\end{eqnarray}
where $\theta_{max}^\omega$ is the angle measured along the perimeter of the right wire. Relatively low losses were assumed (specifically, $\textrm{Im} \left(\frac{\varepsilon_m^\omega}{\varepsilon_{bg}} \right) \ll \left|\frac{\varepsilon_m^\omega}{\varepsilon_{bg}}\right|^2 - 1$). This shows that when the FF approaches the plasmon resonance, the peak position of the linear fields approaches $\theta \to 0$. Conversely, for longer wavelengths the linear electric field, and hence the source term at SH, is pushed more towards the touching point and the attenuation due to the geometric factor increases~\footnote{Similarly, higher losses tend to pull the peak position of $E^\omega_\perp$ away from the touching point; this is indeed intuitive - the absorption limits the near-field enhancement~\cite{Alex_kissing_cyls_NL}.}. In particular, it follows from Eq.~(\ref{eq:th_max}) that since % the higher background shifts the SPP resonance to the red 
the value attained by $|\varepsilon^\omega_m/\varepsilon_{bg}|^2 - 1$ is much smaller for $\varepsilon_{bg} = 12$ than for $\varepsilon_{bg} = 1$% {\bf ??? $\varepsilon^\omega_m(\lambda_{FF}) = - 40.78 + i6.8}$}
, then, $\theta_{max}^\omega$ is found further away from the touching point for $\varepsilon_{bg} = 12$. For a complete map of the $\theta^\omega_{max}$ dependence on the SH wavelength and background permittivity, see Fig.~\ref{fig_poles_soln}(b). % {\bf KNR - explain the problems you found in Alex's expression + the limitation of the current one in a footnote. - had probably a problem near resonance... not so important to us, because we only care about a qualitative description...}

% The maximum value attained at different wavelengths by these near-field quantities and their spatial location is determined by the detuning from the SPP resonance and the absorption\footnote{see~\cite[Eqs.~(39)-(41)]{alex_kissing_NJP}. }. 

%scale=0.40

Recall now that the SH source $J_{z,r}$ is computed by squaring the FF fields and evaluating the tangential derivative along the perimeter of the TW. Thus, faster oscillations gives rise to a stronger SH source. Those are observed for the lower background dielectric case (with a shorter wavelength resonance, see Fig.~\ref{fig_poles_soln}(a)). This phenomenology is the same as reported for increasing frequency and fixed background~\cite{Antonio_kissing_cyls_nonlocal_PRB}: as the frequency gets further away from the surface plasmon resonance, the field oscillation becomes faster. Here, this happens because the growing background permittivity effectively shifts the plasmon resonance to the red. Indeed, Fig.~\ref{fig:sh_diel}(c) and Fig.~\ref{fig:sh_diel}(d) show that the faster oscillations %along the perimeter 
associated with the $\varepsilon_{bg} = 1$ yield a 15-fold stronger source term.

% However, in the proximity of the touching point, the geometric factor takes very small values ($x \approx 0$) such that $x J_{mz}$ is suppressed by the spatial factor $x$ more significant as the wavelength increases.
% Having compared the SH sources, we now compare the total SH response, namely, $H_z^{2\omega}$, for these two cases, 
% The geometric factor counteracts and overpowers a competing effect. 
Yet, the complete SH response (namely, $x J_{z,r}$) is much stronger for the $\varepsilon_{bg} = 12$ case, as seen in Fig.~\ref{fig:sh_diel}(e). % shows that despite the weaker SH source for $\varepsilon_{bg} = 12$, . 
This is the effect of the geometric factor, see Eqs.~(\ref{sh_sol_simp}) and~(\ref{eq:GF}). It acts to suppress the SHG of the $\varepsilon_{bg} = 1$ case more effectively compared with the $\varepsilon_{bg} = 12$ case. %Indeed, we observe that as the source peaks further away from the touching point, the harmful effect of the geometric factor is strongest near the origin ($x \to 0$), is partially neutralized. 
Specifically, the peak value of the SH response is $32.4$ times higher for the $\varepsilon_{bg} = 12$ case, due the combined effect of a $\sim 10$ times stronger maximum value of $|x J_{z,r}|$ (compared to the ratio of $1/15$ for the source only!) and a $\sim 3.1$-fold higher value of $1/\mathcal{P}$.  

Thus, our analysis reveals that having a strong SH source \textit{does not} guarantee a stronger SH response, a potentially counter-intuitively result that stands in contrast to the predictions of approximate models~\cite{Miller_rule,nl_scattering_theory_Bonn}. Instead, it is the spatial distribution of the SH source, i.e., the product $x J_{z,r}$ that determines the near-field SH response. 
% In that sense, it also turns out that the geometric factor has a more significant role compared to the PM factor.

Overall, we have seen that high permittivity values are beneficial for increasing both the PM factor~(\ref{eq:phasematching}) as well as the spatial distribution of the source, $x J_{z,r}$, at least within the regime where Eq.~(\ref{eq:th_max}) is valid. One might then ask what would be the ideal choice of the background permittivity for an optimal SH near-field response. The strongest source would be achieved at the FF surface plasmon resonance, $\varepsilon_{bg} \approx - \textrm{Re}\left[\varepsilon_m^\omega\right]$. Such a choice would also serve to pull the source maximum away from the touching point (see Eq.~\eqref{eq:th_max}) such that the SH near-field becomes stronger. 
% however, while fig. 1 shows an optimal response for values of SH wavelength and permittivity in the middle of the parameter space, fig. 2c shows that optimal response is attained for ... 
On the other hand, a maximal value of $1/|\mathcal{P}|$ corresponds to the SH plasmon resonance, i.e., $\varepsilon_{bg} \approx - \textrm{Re}\left[\varepsilon_m^{2\omega}\right]$. A signature of the contradicting nature of these requirements can be seen in Fig.~\ref{fig_poles_soln}. % only an appropriate choice of $\varepsilon_{bg}$ and $\varepsilon_{bg}^{2\omega}$ (see Fig.~\ref{fig_poles_soln}) can result in SH resonance due to phase-matching condition at that operation wavelength. 

Thus, improving the SH response by taking the geometric factor and phase-matching condition simultaneously into consideration demands the illumination be doubly-resonant, i.e., at both FF and SH. Such a scenario was demonstrated in% the doubly-resonant structure proposed for nanoparticle structures (having discrete spectrum {\bf KNR - you mean non-touching?}) for efficient SHG conversion
~\cite{Taiwanese_double_resonance_films,Italians_double_resonance,Pasha_modal_coupling} and involved non-trivial particle geometries. For the structure studied here, which is characterized by a single, broad resonance, optimal performance can be achieved via a maximally wide resonance bandwidth such that the system gets closer to both resonances. % ({\bf not clear how this satisfies the PM condition or FF resonance condition})

% with (potentially highly) dispersive dielectric background permittivity - the dispersion tunes the SH resonance to $2\omega$. Finding this value and a corresponding realistic implementation is left to a future study. 

\subsection{The far-field response}
\label{sec:farfield}
The discussion so far has focused on the near-fields along the segment of the perimeter closest to the touching point. Turning now to the far-fields, its strength is proportional to the source strength $J_{z,r}$ along the TW perimeter. But as discussed in Appendix~\ref{app:cs} (see Fig.~\ref{CS_1_NumVsAn}), its strength is determined by the values of $J_{z,r}$ further from the touching point, corresponding to the smaller angles displayed in Figs.~\ref{fig:sh_diel}(d)-(e). Thus, the effects of the geometric factor discussed in Section~\ref{sub:near} are essentially irrelevant. 
%In contrast, the fields further away from the touching point are hardly affected by the geometric factor, hence, they are simply proportional to the source $J_{z,r}$, see Figs.~\ref{fig:sh_diel}(d)-(e).
Instead, it is the decay length of the fields away from the touching point that determines the far-field response. This decay is slower for the $\varepsilon_{bg} = 12$ case (see Fig.~\ref{fig:sh_diel}(b) and~(d)) so that indeed, the far-field response in that case os much stronger. In particular, the source term for the $\varepsilon_{bg} = 12$ case at angles away from the touching point is about 5 orders of magnitude greater than for $\varepsilon_{bg} = 1$! This originates from the fact that the FF electric field in the source term is $\sim 2$ orders of magnitude higher for the $\varepsilon_{bg} = 12$ case (see Fig.~\ref{fig:sh_diel}(b)) due to the proportionality to the background permittivity and to $\alpha^\omega$ (which represents field confinement and the propagation constant), see Eq.~(\ref{sh_sol_simp}). Both of these are higher by an order of magnitude for the $\varepsilon_{bg} = 12$ case. The differentiation of Eq.~(\ref{eq:jmz}) extracts a further factor of $\alpha^\omega$, hence adding another order of magnitude. % ({\bf I see 4 orders in Figure 5}).. I see 5.. and anyhow, we write about 5.. 
This result was obtained in~\cite{alex_kissing_NJP}; again, its significance was not emphasized, yet, its origins were. Indeed, only the ratio between permittivities appears in quasi-static expressions in~\cite{alex_kissing_NJP} such that increasing the permittivity in the background is equivalent to reducing the absolute value of the metallic one. As a result, the highest confinement take place at large, negative values of the metal permittivity (i.e., at long frequencies of a Drude metal) or for high permittivity background. As explained in Section~\ref{sec:spectrum} below, this behaviour also explains the spectrum of the scattered SH power.

% I believe it derives from the amplitude of the linear solution that appears in the sources. The increase in amplitude depends on $const/(x+iy)^2$ when the "$const$" contains $\alpha$ and $\varepsilon_{bg}$ and is much larger for higher background permittivity.

As a result, the scattered power (which scales with the square of the fields) is about 10 orders of magnitude greater for $\varepsilon_{bg} = 12$ than for $\varepsilon_{bg} = 1$, see Fig.~\ref{fig:sh_diel}(f). This is the second main result of the current work. It is particularly remarkable in comparison to the single wire, where a change in $\varepsilon_{bg}$ improves the scattered power by ``only'' 2 orders of magnitude (not shown). Sections~\ref{sub:near} and~\ref{sec:farfield} have demonstrated that the $\varepsilon_{bg} = 12$ case exhibits superior behaviour both in the SH near- and far-field regimes of the TWs.

% , see Figs.~\ref{TWsVsSingle}. % This data is presented in \url{mono-dimer-cs-comparison_1bg} and \url{fig:TWsVsSingle12}}} and~\ref{fig:TWsVsSingle12} below. 
%\begin{figure}[H]
%\centering
%\includegraphics[width=0.55\textwidth]{F10_CS_12Vs1_1.png}
%\caption{The numeric solution for the SH scattering cross-sections of TWs for $\varepsilon_{bg} = 12$ (dots) and $\varepsilon_{bg} = 1$ (pluses). } \label{TWs12VsTWs1}
%\end{figure} 

\section{The SH spectral response of the identical touching wires}\label{sec:spectrum}

From the analysis of Section~\ref{sub:near}, it became clear that one way to optimize the SHG efficiency of the TWs is to increase the bandwidth of its spectral response. In fact, the unusually large bandwidth of the linear response of the TWs was also our original motivation to study SHG from the TWs~\cite{alex_kissing_NJP,Alex_kissing_cyls_NL}. Moreover, %why moreover? is this not the explanation for the previous sentence?}) - no.. the following discusses a unique phenomenon associated with nonlinear wave mixing whereas large bandwidth of the linear response has a different origin.
a further increase in the SH bandwidth was expected due to the effects of auto-resonance, also known as adiabatic frequency conversion~\cite{Porat_review}. This was demonstrated in tapered dielectric waveguides~\cite{Yaaqobi-friedland}, and relies on sweeping the phase-matching parameter (analogous to $\mathcal{P}$) adiabatically through zero along the propagation direction instead of achieving the exact phase-matching within the entire nonlinear medium. Since the tapered waveguide is mimicked by the TW geometry near the touching point for the modes circulating around the particle perimeter (e.g., Fig.~\ref{fig:TWs}), a similar enhancement was expected.
% This behavior would complement the large field enhancement occurring near the singular point, which was already used to demonstrate efficient nonlinear frequency conversion~\cite{koreans-HHG-taper,Polman_taper_Erbium}.
Unfortunately, as we shall see below, the SH response of the TWs did not meet such expectations. Instead, the response turned out to be narrowband, as it typically is. % ({\bf should refer to a figure. is 5(d) correct?})

\begin{figure}[H]
\begin{center}
\includegraphics[width=1\linewidth]{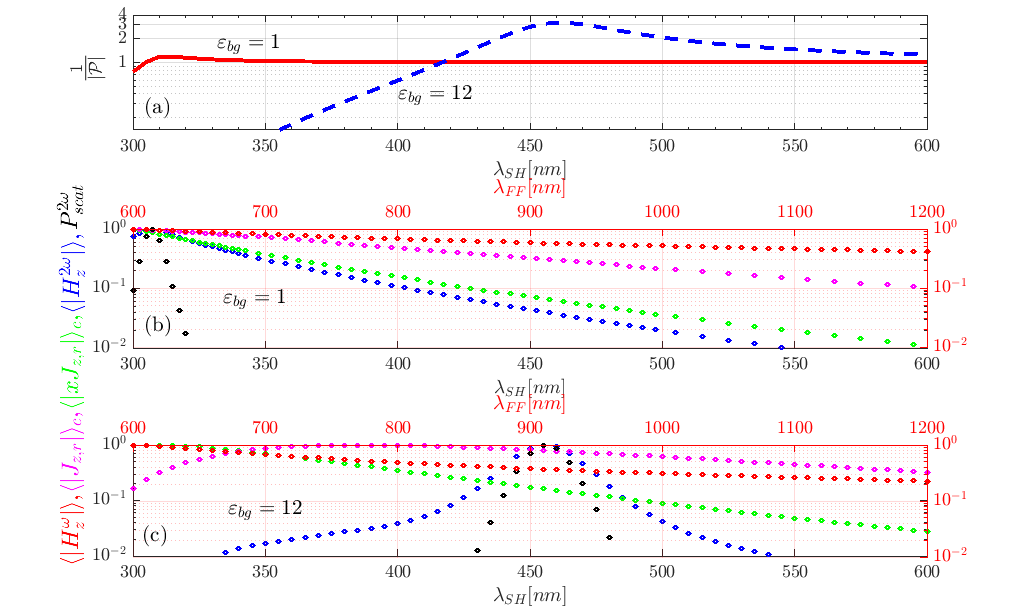}
\caption{(Color online) %{\bf SE - can you try to make each plot taller? maybe using subplot(x,y,'align')? or any other matlab trick or some external software? YS: This is the maximum in MATLAB, i tried also with Gimp and it looks bad, please see \url{SpectralAnalysisEpsilon12_P_.png} } 
Spectral responses of the SH near-field of the identical TW system. (a) The PM factor $1/|\mathcal{P}|$~(\ref{eq:phasematching}) as a function of SH wavelength for $\varepsilon_{bg} = 1$ (solid red line) and $\varepsilon_{bg} = 12$ (dashed blue line). Normalized average of the near-field quantities along the perimeter of the TW, namely, $|H_z^\omega|$ (red dots)$|J_{z,r}|$ (magenta dots), $|xJ_{z,r}|$  (green dots),  $|H_z^{2\omega}|$ (blue dots) and $P^{2\omega}_{scat}$ (black dots) for (b) $\varepsilon_{bg} = 1$, (c) $\varepsilon_{bg} = 12$. Each curve is normalized by its maximum value in the spectral range of interest. % ({\bf I still question why (b) and (c) are so complex. Only the black and blue curves are referenced in the text.})
% It would also make it easier to interpret. I think the busyness of (b) and (c) along with their aspect ratios make them difficult to interpret. Perhaps expand the 2 plots into 4 plots with only 3 lines each.}) - plots are indeed busy, but I think splitting them will make things less clear. these are plots that require time to digest, indeed...
} \label{fig:sp_res} 
\end{center}
\end{figure}

To understand why, we first consider the spectral behavior of the SH source. It is obtained by squaring the FF field, so that the spectral bandwidth of the SH source is necessarily narrower than the linear response (compare the red and magenta curves in Fig.~\ref{fig:sp_res}(b); the effect is less pronounced in Fig.~\ref{fig:sp_res}(c)). %({\bf Does this really change $\Delta\omega/\omega$?}) %this mixes spectrum and space... SE - what did you want to say here?\textcolor{blue}{i didn't write this... He tried to say that the spectral bandwidth of the SH source is reduced compared the spectral bandwidth of the linear case(FF fields)}. %  show the linear and SH source's spectral response for $\varepsilon_{bg}^{\omega,~2\omega}=1$ and $12$, respectively, confirming our prediction. 
% We now focus on the resulting SH spectral response. As predicted by the analytical solution, the product of the spectral behavior of the phase-mode matching and SH source gives us the SH spectral response. This implies that the SH response should roughly scale as the SH source as the phase-mode-matching factor is unity in long wavelength limit. However, in contrast to this prediction, the SH spectral response (obtained by averaging $H^{2\omega}_z$ over the TW) doesn't scale as the SH source's response as seen in Fig.~\ref{sp_res}(b)-(c). This na{\"i}ve prediction is erroneous for the following reasons: The complete analytical solution [see Eq.~\eqref{sh_sol_simp}] also includes the geometric factor $x$. 
Another limitation on the bandwidth is imposed by the geometric factor $x$ (see green curves in Figs.~\ref{fig:sp_res}(b)-(c)). Indeed, one might na{\"i}vely conclude that that this term is insignificant in determining the spectral response as it is only a spatial factor that remains unchanged regardless of wavelength, but Eq.~(\ref{eq:th_max}) and Fig.~\ref{fig_poles_soln}(b) show that the opposite is true. The geometric factor is less harmful near the SH resonance; this is due to the wavelength dependence of the maximal field position, see the discussion surrounding \eqref{eq:th_max}. Thus, the geometric factor plays a significant role, and determines not only the near-field enhancement (see Fig.~\ref{fig:sh_diel}(e)), but also the spectral bandwidth of the SH near-fields~\footnote{It is worth noting that the additional factor $\sim \omega \sinh 2 \alpha^\omega / \alpha^\omega$ also contributes to the spectral narrowing. }. In contrast, Fig.~\ref{fig:sp_res}(a) shows that the peak position is determined by the PM factor~(\ref{eq:phasematching}).

Finally, we observe that both the far-field and near-field responses peak at the same wavelength, yet the bandwidth of the far-field response is significantly narrower (compare black and blue curves in Figs.~\ref{fig:sp_res}(b)-(c)). This difference can be understood by recalling that the spectral width of far-field resonances is governed by radiative damping, whereas absorption losses dominate the near-field. The latter are much faster than the former for small particles roughly a few tens of nanometers in radius like the ones considered in our study~\cite{nano-optics-book}. This well-known effects provides further spectral narrowing on top of that caused by the geometric factor. % ({\bf However, the far-fields are not affected by the geometric factor, so why should it be narrower than the near-fields?})% ({\bf Is this because greater losses causes a broader peak? But then the peak value is also lower, and this information is lost when the curves are normalized.})

% The value of the scattering cross-section should depend strongly on the TWs size {\bf well - already mentioned above...} and the background permittivity. 
% The bandwidth of $\sigma_{scat}$ is, however, narrower compared with that of the (averaged) near-field response. Indeed, since the latter relates to the absorption cross-section, this behaviour is typical for any small NP (in the quasi-static limit)~\cite{?} {\bf Antonio - can you look for a relevant reference? any decent textbook would be fine, no?} {\bf is anything like that was discussed by Paloma / Fan Yang?}

These results show why, contrary to expectations, the bandwidth of the SH response is much more limited than the linear response. In retrospect, the TO approach reveals that although the geometry we study resembles a tapered waveguide, it is fundamentally based on the flat slab geometry, for which no auto-resonance response is expected.

% {\bf I think that the text is hard to follow for reader, reference to physical magnitudes barely defined in the text - Antonio - see if any better now. please make suggestions/corrections that will yield further clarity}

\section{comparison to previous work and outlook}
The goal of the current work was to highlight the utility of TO as a tool for the interpretation of the wave physics  of nonlinear optical wave interactions in plasmonic nanostructures, and the optimization of its performance. This was demonstrated through a thorough analysis of a well-studied singular plasmonic structure---the touching wires (TWs)---using the simplest possible solid-state model for the second-order optical response of metals.% This revealed various insights of general relevance - {\bf permittivity - 1 vs. 12 (can we explain this further - pulling field out of touching point - opposite to linear motivation). ... }

% {\bf get feedback from Lippitz and Haim for the analysis below...} 
The combined near- and far-field analyses provided in this work enables comparison with some commonly used theoretical approaches employed for SHG in nanostructures. To perform this comparison, we first recall that the exact formal solution for the SHG problem from an arbitrary scatterer is simply the convolution of the SH current with the Green's tensor ($\tensor{G}$), which represents the response of the structure to a point dipole source (i.e., the spatial impulse response of the scatterer); it is generically a complicated function of space and frequency. Our analysis of the TWs showed that some aspects of the far-field response, such as the permittivities that yield stronger responses, can be correctly predicted by considering only the nonlinear polarization (i.e., the source, $J_z$), Fig.~\ref{fig:sh_diel}(d) and~(f). % ({\bf There is still a logical leap between the preceding and following sentences. There is a sudden and unstated transition between the importance of the full Green's tensor and a constant Green's tensor. It is perhaps because the previous sentence is vague and incomplete.}) 
This approach is essentially Miller's rule, which is applicable to bulk materials that exhibit a relatively weak spectral sensitivity, and involves fairly uniform fields (i.e., no complex scattering due to subwavelength structuring). In this case, the Green's tensor exhibits weak variations in space and frequency, such that the general solution is indeed simply proportional to the SH source. This polarization-based model was also shown to be successful when applied to systems far from resonance, as e.g., in~\cite{Lippitz_Giessen_THG} or those that have a relatively broad resonance, as in the current case (see Figs.~\ref{fig_poles_soln}(a) and~\ref{fig:sp_res}(a)). 

However, our solution shows that the {\em near}-field exhibits a more complicated form. It includes considerations such as proximity to the SH resonance (which we refer to as phase matching, PM) and symmetry of the source and modes (via the more quantitative concept of mode matching, MM). These considerations arise naturally in the so-called nonlinear scattering / effective susceptibility model~\cite{nl_scattering_theory_Bonn,Haim_overlap} % ({\bf Is this method being introduced for purposes of comparison with TO? If so, say so.}) 
whereby the product of the SH polarization with the linear field is spatially averaged. This approach was shown %({\bf here or elsewhere?}) - elsewhere, hence, reference at the end of sentence
to predict correctly various additional quantitative aspects of the solution (e.g., the optimal structural asymmetry~\cite{Haim_overlap}). % {\bf... and probably also the right spectral position, but not the bandwidth?)..}.

However, that approach does not accurately handle the interplay between the SH source ($\sim J_{z,r}$) and SH response ($\sim \tensor{G}$), in that it does not rely on the exact Green's tensor of the structure~\cite{GENOME}, instead relying on the profile of an incident plane wave. Nor does it correctly consider the convolution between these two quantities, instead relying on their product. The former approximation may yield discrepancies in the spatial and spectral responses while the latter approximation can lead to somewhat different interference effects compared with straightforward averaging~\cite{nl_scattering_theory_Bonn}. It is clear that these effects did not prevent the good agreement between analysis and measurement in the far-field response reported by~\cite{Haim_overlap}, but it may not correctly predict the near-field~\cite{nl_scattering_theory_Bonn}. Since our exact solution is free of the approximations associated with the nonlinear scattering theory of~\cite{nl_scattering_theory_Bonn}, it indeed predicts a near-field behaviour that differs from the far-field, as seen in Figs.~\ref{fig:TWsVsSingle12}-\ref{fig:sh_diel}. This is due to the appearance of the geometric factor, which modifies the near-field enhancement (see Fig.~\ref{fig:sh_diel}(e)), and controls the overall bandwidth of the spectral response, see Fig.~\ref{fig:sp_res}. All the above shows that polarization-based models such as Miller's rule or simplified models such as the nonlinear scattering model / effective susceptibility model~\cite{nl_scattering_theory_Bonn,Haim_overlap,Lippitz_nl_scattering_model} should be used with caution in structured nanophotonic systems. Nevertheless, these simpler models are applicable to wide class of systems, while it is not clear how the additional insights revealed in our work apply to other structures.

In view of the above, we emphasize that the discussion in the current work should not be construed as anything more than a qualitative study. Indeed, various effects are expected to modify the quantitative results shown here, such as nonlocality~\cite{Antonio_kissing_cyls_nonlocal_PRL,Antonio_kissing_cyls_nonlocal_PRB}, dispersion of the second-order response (which is not yet fully-understood, in general), interband transitions~\footnote{Those occur for the Ag used here on wavelengths shorter than $\approx 350$nm; they were intentionally neglected in order to allow focusing on the wave aspects of the problem. These effects would naturally increase the absorption and limit the near-field enhancement and the scattering.} etc. These reduce the near-field enhancement, for example. Conversely, employing a more realistic permittivity than the one we used may have the opposite effect. Nevertheless, the analytic solution and insights obtained in this work can lead towards various ways of improving the response quantitatively, for example, to the material dispersion or anisotropy of the background medium necessary for the optimal response. More significant improvements may be achieved if the background also has a second-order optical response; this is particularly likely for the high permittivity background which was found to yield superior performance in this study. In that sense, it would be interesting to build on recent work on SHG from such systems~\cite{Vincenti:2011wn,Scalora:2012gw,de_Angelis_AlGaAs_nanodimers,de_Angelis_LinBo}, and to study the relative importance of the SHG from each material.

In addition to its insights on the TWs, the current study can be extended to a variety of additional plasmonic nanostructures. Specifically, one can exploit the results for dipole sources above a metal layer in~\cite{Zayats_dipoles_near_surface_2019} to study the field distribution due to illumination by electric fields of different polarizations as well as to study other configurations such as the crescent structure~\cite{Alex_crescent}, non-touching wires~\cite{Alex_hybridization}, blunt geometries~\cite{Yu_bluntness}, periodic arrays, 2D materials~\cite{Fan_metasurfaces,Paloma-review-TO}, and various 3D structures~\cite{Antonio_kissing_spheres_PRL,TO_vdw_PNAS} etc. Future steps may include further optimization of the performance, e.g., by exploiting the greater susceptibilities reported within hetero-dimers~\cite{Niv_SHG_2018,Niv-Schvartzman-Sivan} or even to treat more complicated nonlinear wave interactions such as sum frequency generation~\cite{Fan_THz_MDM,Che2016}, third-harmonic generation~\cite{Haim_overlap,Lippitz_Giessen_THG,Lippitz_Giessen_THG2}, THz generation~\cite{Fan_THz_MDM,Ellenbogen-Minerbi} and four wave mixing~\cite{Oulton_4_wave_mix}.

Finally, our study will hopefully motivate an experimental test of our predictions. Indeed, many aspects of the linear response of TWs revealed by earlier analysis were already experimentally demonstrated using thin discs (in the THz regime), see~\cite{Hanham_AdvMat,Lei_sphere_plane}.

% when reducing the losses consistently through a Drude model, the peak value grows but the field in the region away from the touching point drops slightly but remains at the same place - so just sharpening? See \texttt{losses geom FF and SH.png} - SE/KNR - where are these plots? different from fig. 4?

% what does the move between 1 and 12 do to the losses (resonance quality etc.) - apparently, not much

% can we compare to  "loss as a route for transparency"; to STIRAP? 

\section{Acknowledgements} 
KNR and YS were supported by Israel Science Foundation (ISF) grant (899/16). AIFD acknowledges funding from the Spanish MICIU under contract RTI2018-099737-B-I00, and from the BBVA Foundation through a 2019 Leonardo Grant for Researchers and Cultural Creators.

\appendix

\section{The analytic solution for the SH magnetic field near the touching wires}\label{app:PRB}

In~\cite{Reddy_SHG_TO}, we calculated the second harmonic (SH) field generated by a pair of (identical) touching metal wires (TWs, Fig.~\ref{fig:TWs}) assuming the simplest second-order nonlinear polarization of the metal, i.e., assuming it is dominated by a perpendicular {\em surface} source. As explained in~\cite{Kauranen_perp_perp_perp,Reddy_Sivan_SHG_BCs,Reddy_SHG_TO}, this is a reasonable generic (even if not complete) description of the second-order nonlinear optical response of metals; indeed, {\em bulk} polarization can also be accounted for by mapping it to a surface polarization. We showed that a highly accurate analytic solution can be obtained in three steps. First, we transformed the SH source~\footnote{Note that this is far simpler than transforming each of the constituents of the polarization source - the $\chi^{(2)}$ tensor, and each of the FF fields. } (calculated within the quasi-static approximation~\cite{Alex_kissing_cyls_NL,alex_kissing_NJP}) and boundary conditions to the slab frame (see Fig.~\ref{fig:TWs}) using a conformal inversion transformation ~\cite{Alex_kissing_cyls_NL}. Then, by assuming that the amplitude of the excited guided wave in the slab geometry is slowly varying along $\tilde{y}$, we imposed the boundary conditions appropriate for the surface source (namely, continuity of the magnetic field and proper discontinuity of the parallel electric field~\cite{Reddy_Sivan_SHG_BCs})~\footnote{This approach proved to be simpler than the WKB approximation employed in~\cite{Antonio_kissing_cyls_nonlocal_PRL,Antonio_kissing_cyls_nonlocal_PRB} for the linear non-local response. }. Finally, we transformed back to the TW frame and obtained % {\bf I would not give any details here, it is more confusing than clarifying, unless you make proper definitions and explain in more detail how Eq. (1) was obtained. I would give more details about Fig. 1, or I would remove it. I think it may be against our interest to include it. It has been seen too many times, I think.} 
\begin{widetext}
%\begin{footnotesize}
\begin{eqnarray}
H_z^{2\omega} &=& \frac{-i \omega \varepsilon_0}{2\alpha^\omega \mathcal{P}(\omega)} \mathcal{R}\left(\tau_x,\tau_y\right) \mathcal{G}(x,y) \exp\left(\frac{4 i a \alpha^\omega y}{x^2 + y^2}\right) \nonumber \\
&\times&
\begin{cases}
\sinh\left(\frac{4 a \alpha^\omega x}{x^2 + y^2}\right), \quad x^2 + y^2 + 2|a| x > 0, \\
e^{2\alpha^\omega} \sinh 2\alpha^\omega \exp\left(\frac{- 4 a \alpha^\omega x}{x^2 + y^2}\right),\quad x^2 + y^2 + 2|a|x < 0.
\end{cases} \label{eq:ShMagneticField} 
\end{eqnarray}
%\end{footnotesize}
\end{widetext}
Here, $x$ and $y$ are the real space coordinates, $\alpha^\omega = \frac{1}{2} \ln\left(\frac{\varepsilon^\omega_m - \varepsilon^\omega_{bg}}{\varepsilon^\omega_m + \varepsilon^\omega_{bg}}\right)$ is the dimensionless~\footnote{$\alpha^\omega = |k|d$ where $k$ is the spatial Fourier transform coordinate; in the quasistatic limit, it represents the propagation constant as well as the mode transverse width. } propagation constant at the fundamental frequency $\omega$,~$\varepsilon^\omega_{bg}$, $\varepsilon^\omega_m$ and $\varepsilon_0$ are the background, metal and vacuum permittivities, respectively, $a$ is the wire radius, and $\mathcal{P}$~(\ref{eq:phasematching}) is the so-called Phase Matching (PM) factor. It represents the pole of the dispersion relation of the antisymmetric mode. It is maximal at resonance, i.e., when the SH frequency is tuned to the frequency of that mode~\cite{Reddy_SHG_TO}~\footnote{As explained in~\cite{Reddy_SHG_TO}, $\mathcal{P}$ differs from the potentially more familiar PM factor, see e.g.~\cite{Boyd-book} (which is $\textrm{sinc}[(\alpha^{2\omega} - 2\alpha^\omega)d]$); they vanish together, but $\mathcal{P}$ is more informative as it includes information about the profile of additional modes (of the same symmetry). }. $\mathcal{R}$ is related to $J_{z,r}(x,y)$, the ($z$-directed) surface magnetic current generated by the nonlinear polarization on the right wire. It is evaluated via the analytic solution for the fundamental frequency (FF) fields, $E_\perp^\omega$, provided in~\cite{alex_kissing_NJP} and is itself given by Eq.~(\ref{eq:jmz}) such that $\mathcal{R}$ is defined via
\begin{eqnarray}
J_{z,r}(x,y) &=& \mathcal{R}(\tau_x,\tau_y)~\exp{ \left( \frac{4 i a \alpha^\omega |y|}{x^2 + y^2}\right)} ~\delta(x^2 + y^2 - 2ax). \label{eq:Decpomposedjmz}
\end{eqnarray}
Here, %{\bf SE - where did the taus go missing? shouldn't the arguments of $R$ be those taus? ANS: Taus are obtained from the transformation and BC solution. Here we present the sources along the TWs contour before the transformation, however if you inject the contour coordinates (delta function values, $x^2 + y^2 - 2ax$), you get $x$,$y$ instead the taus } 
the coordinates $\tau_x$ and $\tau_y$ are defined as 
\begin{equation}
(\tau_x,\tau_y) = \left(\frac{1/(2a)}{1/(4a^2) + y^2/(x^2 + y^2)^2},\frac{y/(x^2 + y^2)}{1/(4a^2) + y^2/(x^2 + y^2)^2}\right). \label{eq:tau}
\end{equation}
These coordinates map points in the domain outside the TWs to the TW perimeter, see Fig.~\ref{CS_1_NumVsAn}(a). Finally, the factor % {\bf Note that a is not defined either. I do not see clearly how to present this.}
\begin{equation}
\mathcal{G}(x,y) = \frac{4a(x^2 + y^2)^2}{4a^2y^2 + (x^2 + y^2)^2}. \label{eq:GF}
\end{equation}
is referred to as the geometric factor. It originates from a transformation of the generalized boundary condition (which incorporates the discontinuity in the tangential electric field) to the slab geometry~\cite{Reddy_SHG_TO} as well as from the complex non-uniform distributed nature of the source. Along the TW perimeter, $\mathcal{G}$ reduces to simply $x$, as shown in~\cite{Reddy_SHG_TO}; this is used in Eq.~(\ref{sh_sol_simp}).

\section{Scattered power calculations}\label{app:cs}

Scattering cross-sections are probably the most accessible observable in experimental studies of small nanoparticles. However, they vanish in the quasi-static limit, which was the limit used to calculate the linear electric fields~\cite{Alex_kissing_cyls_NL,alex_kissing_NJP}. Previous works (e.g.,~\cite{Alex_beyond_quasistatics}) went beyond this limit by computing the linear absorption and scattering cross-sections of the TWs based on the dipole moment induced by the incident fields. As in the case of the near-field enhancement, the TWs exhibited a scattering cross-section which is spectrally much broader compared to the single wire. % especially for identical TWs which shows the most broadband response, in the near-infrared and visible regimes, 
% see Fig.~\ref{TWsVsSingle}(a) and~\cite{Alex_beyond_quasistatics}. {\bf This was done by defining a net dipole of the TWs in the transformed geometry~\cite{alex_kissing_NJP}. Its product with the dyadic Green’s function gave the scattered dipole that is directly proportional to the electric field in the slab frame. Consequently, they were able to define a fictitious particle which absorbs the radiation in the slab frame. Finally, since the absorption power is well known, the scattering cross-section was found without any integration by injection the scattered dipole and the slab electric field into the known absorption power expression. they calculated the absorption power of the fictitious particle in the slab which can be represented by the scattered power in the TWs frame. They used the particle absorption power equation $P_{abs} = - \frac{\omega}{2} Im(p_{dipole}*E^*_0)$. This solution process can not be performed in our case since we cannot define a net dipole as in~\cite{Alex_beyond_quasistatics}. }
Hence, it is interesting to see if a large bandwidth is obtained also for the SH response. 

% branch cut contributes to the far field. this is why initially Alex did not compute it. in the second stage, he calculated it as described above.

\begin{figure}[H]
\centering
\includegraphics[width=1\linewidth]{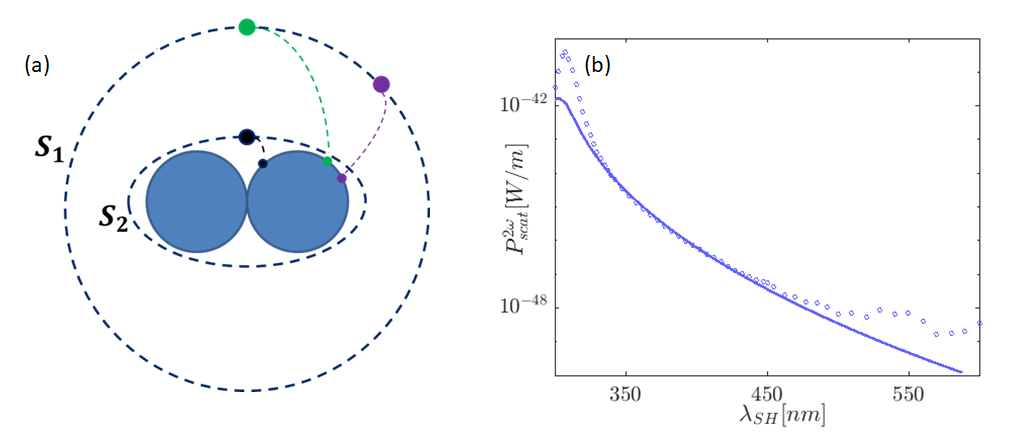}
\caption{(Color online) (a) Two surfaces $\mathcal{S}_1$ and $\mathcal{S}_2$ surrounding the TWs. $\tau_x$ and $\tau_y$ map each point on these surfaces to the TW perimeter. For $\mathcal{S}_1$, with a radius of 22nm, each point maps to points relatively far from the touching point. Conversely, for $\mathcal{S}_2$ which is closer to the TW structure, each point maps closer to the touching point. (b) The numeric (dots) and the analytic (blue solid line) solutions for the SH scattered power of TWs for $\varepsilon_{bg} = 1$.} \label{CS_1_NumVsAn}
\end{figure}

Unfortunately, the approach of~\cite{Alex_beyond_quasistatics} cannot be applied in our case, as a total SH dipole moment cannot be defined for the nanostructure at the SH. Instead, we exploit the inherent near-to-far field mapping which is manifested by $\mathcal{R}(\tau_x,\tau_y)$~(\ref{eq:Decpomposedjmz}). As noted, this factor map points in the domain outside the TWs to the TW perimeter, see Fig.~\ref{CS_1_NumVsAn}(a), in such a way that % allows us to write $\mathcal{R}(\tau_x,\tau_y)$ as 
% \begin{eqnarray}
% \mathcal{R}(\tau_x,\tau_y) \exp\left(\frac{i 4a \alpha^\omega|\tau_y|}{\tau_x^2 + \tau_y^2}\right) = J_{r,z}(\tau_x,\tau_y) = \chi^{(2)}_{S,\perp\perp\perp}~\partial_\parallel[E_\perp^\omega(\tau_x,\tau_y) E_\perp^{\omega}(\tau_x,\tau_y)], \label{eq:R}  
% \end{eqnarray}
%In addition, the field compression is higher as it get closer to the touching point {\bf so what? why relevant?}. % Therefore, the ability to calculate these cross-sections depends on the space mesh resolution. 
more distant contours correspond to points further away from the touching point. In fact, since the field varies more slowly in these regions compared to the regions closer to the touching point, it is easier to calculate the scattered power rather than the absorbed power (which requires an integration over the rapidly varying near-fields). Peculiarly, this magnetic-field-based calculation somehow transcends the quasistatic limit, thus, providing a non-zero prediction for the scattering cross-section. 

Thus, we now proceed past the near-field calculation of~\cite{Reddy_SHG_TO} and compute analytically and numerically the SH scattered power from the TWs, $P^{2\omega}_{scat}$,
% this is easier than calculating the absorbed power - because the latter requires a 2D integration that includes the highly oscillatory fields near the touching point, whereas the scattered power requires just a line integral over fields that vary more slowly.
given by
\begin{eqnarray}
P^{2\omega}_{scat} = \frac{1}{2} Re \int_{\mathcal{S}} E_\phi^{2\omega} \left[H^{2\omega}_z\right]^* d\mathcal{S}. \label{eq:Power}
\end{eqnarray}
This is the scattered power that passes through the surface $\mathcal{S}$, where $\hat{n}$ represents the normal to the surface, see Fig.~\ref{fig:TWs}(b). 

To calculate the SH magnetic field~(\ref{eq:ShMagneticField}), we have to evaluate $J_{r,z}(\tau_x,\tau_y)$~(\ref{eq:Decpomposedjmz}); this requires computing the analytic expression for the parallel derivative of the FF electric field $E_\perp^\omega(\tau_x,\tau_y)$. It is given by
\begin{eqnarray}
\partial_\parallel[E_\perp^\omega(\tau_x,\tau_y)] = \mathcal{M} + \mathcal{N}, \nn \\
\end{eqnarray}
where
\begin{widetext}
\begin{footnotesize}
\begin{eqnarray}
\mathcal{M} = \frac{i y \pi \alpha^{\omega}a E_{0x}^{\omega}}{\varepsilon^{\omega}_m + 1} \left[ \frac{- i 2 a y^2 - i\left(x - a\right)^2 \left(2 x - \alpha^\omega a\right) + \frac{1}{2} a |y|\left(4 a - \frac{6 y^2}{a} - 2 \alpha^\omega a \left(\frac{x - a}{a}\right)\right) + |y|^3}{|y|\left(x + i|y|\right)^4} \right]\exp\left(\frac{-\alpha^\omega a}{x + i|y|}\right) , \nn \\
\label{E_derivative}
\end{eqnarray}
\end{footnotesize}
and
\begin{footnotesize}
\begin{eqnarray}
%\mathcal{N} = \frac{(x-a)\pi\alpha^{\omega}a}{\varepsilon^\omega_m + 1} \exp\left(\frac{-\alpha^\omega a}{x + i|y|}\right) \frac{- \textrm{sign}\left[y\right]}{\left(x + i|y|\right)^2} - \frac{y \pi \alpha^\omega a \exp\left(\frac{-\alpha^\omega a}{x + i|y|}\right) \left[|y| - i\left(x - a\right)\right] \left(i2|y| + 2x -\alpha^\omega a\right)}{(\varepsilon^{\omega}_m + 1)\left(x + i|y|\right)^4}. \nn \\
\mathcal{N}=\frac{- \textrm{sign}\left[y\right]\pi\alpha^{\omega}aE_{0x}^{\omega}}{\varepsilon^\omega_m + 1}\left[ \frac{(x-a)}{\left(x + i|y|\right)^2} - \frac{y  \left[|y| - i\left(x - a\right)\right] \left(i2|y| + 2x -\alpha^\omega a\right)}{\left(x + i|y|\right)^4}\right] \exp\left(\frac{-\alpha^\omega a}{x + i|y|}\right). \nn
\end{eqnarray}
\end{footnotesize}
\end{widetext}
where $E_{0x}^{\omega}$ is amplitude of the incident plane wave along $x$ direction. 
Now that we have all the expressions needed for the evaluation of the magnetic field~(\ref{eq:ShMagneticField}), we can analytically calculate the SH parallel electric field, % $E_\parallel^{2\omega}(\tau_x,\tau_y)$, 
using
\begin{eqnarray}
% \vec{E}(r = R) = \frac{i}{\omega\varepsilon_0\varepsilon^{\omega}_{m}}\vec{\nabla}\times H_z \rightarrow 
E_\phi^{2\omega}(r) = \frac{i}{2 \omega \varepsilon_0 \varepsilon^{2\omega}_m} \frac{\partial H_z^{2\omega}}{\partial r} \label{radialderiv}
\end{eqnarray}
from the analytic expression for the magnetic field~(\ref{eq:ShMagneticField}). Here, $r = \sqrt{x^2 + y^2}$ is the radial coordinate along which the differentiation is performed. % The derivative adds another coordinate to the data calculation since all the expressions in this section depend the parallel variable with fixed radial value. This addition makes the analytic calculation long and tedious. 

In order to validate the solution described above, we support our analytic results by direct numeric simulations using COMSOL Multiphysics (V 3.5a), see Appendix E in~\cite{Reddy_SHG_TO}. In fact, the numeric calculation of the derivatives in Eq.~(\ref{radialderiv}) required even higher resolution compared to the calculations in~\cite{Reddy_SHG_TO}.

We observe good agreement between the analytic and numeric calculations for most of the spectral regime studied, with the exception of the lower frequency range, see Fig~.\ref{CS_1_NumVsAn}. As we saw before (Eq.~(\ref{eq:th_max})), the fields peak closer to the touching point when material losses are lower, and are also more concentrated at lower frequencies, see Fig.~\ref{fig:sh_diel}(a). Thus, capturing this effect in numerical simulations is naturally harder. Indeed, the numeric results for the Drude loss parameter $\gamma = 0.2$ deviate from the analytic solution in the long wavelength regime; better agreement was obtained in this regime for higher absorption (not shown). % Due to this discrepancy we changed the losses coefficient to higher value, $\gamma_{loss} = 0.4$ and we got smooth behaviour, see Fig.~\ref{CsAnalyticVsNumeric}.

\bibliography{my_bib}

\end{document}